\documentclass[12pt]{article}

\usepackage{jheppub}
\makeatletter
\def\@fpheader{\relax}
\makeatother

\usepackage{caption}
\usepackage{subcaption}
\usepackage{amsmath,amssymb,amsfonts,graphicx,slashed,color,
amsthm,mathtools,upgreek}

\title{Deep Learning the Hyperbolic Volume of a Knot} 

\author[a, b]{Vishnu Jejjala}
\author[b]{\!, Arjun Kar}
\author[b]{\!, Onkar Parrikar}

\affiliation[\,a]{Mandelstam Institute for Theoretical Physics, School of Physics, NITheP, and CoE-MaSS, \\ University of the Witwatersrand, Johannesburg, WITS 2050, South Africa}
\affiliation[\,b]{David Rittenhouse Laboratory, University of Pennsylvania,\\
209 S 33rd Street, Philadelphia, PA 19104, USA}

\emailAdd{vishnu@neo.phys.wits.ac.za, arjunkar@sas.upenn.edu, parrikar@sas.upenn.edu}

\abstract{
An important conjecture in knot theory relates the large-$N$, double scaling limit of the colored Jones polynomial $J_{K,N}(q)$ of a knot $K$ to the hyperbolic volume of the knot complement, $\text{Vol}(K)$.
A less studied question is whether $\text{Vol}(K)$ can be recovered directly from the original Jones polynomial ($N = 2$).
In this report we use a deep neural network to approximate $\text{Vol}(K)$ from the Jones polynomial.
Our network is robust and correctly predicts the volume with $97.6\%$ accuracy when training on $10\%$ of the data.
This points to the existence of a more direct connection between the hyperbolic volume and the Jones polynomial.
}

\date{}

\newcommand\eref[1]{(\ref{#1})}

\begin{document} 

\parskip=10pt

\maketitle 

\section{Introduction}
Identifying patterns in data enables us to formulate questions that can lead to exact results.
Since many of these patterns are subtle, machine learning has emerged as a useful tool in discovering these relationships.
In this work, we apply this idea to invariants in knot theory.

A knot is an embedding of a circle in the $3$-sphere $S^3$.
These objects play important roles in a wide range of fields including particle physics, statistical mechanics, molecular biology, chemistry, sailing, and art~\cite{Witten1989,Jones1990,Sumners1995,Horner2016,Ashley1944,Jablan2012}.
Figure~\ref{fig:knots} depicts several well known knots.
\newcommand\one{From left to right: the unknot, trefoil knot, figure-eight knot, and cinquefoil knot. 
When strands of the knot cross, the diagram keeps track of which strand is on top and which strand is on the bottom, so the diagram captures all information about the $3$-dimensional embedding.
For the figure-eight knot, the Jones polynomial in our conventions is $J_{\text{figure-eight}}(q) = q^{-2}- q^{-1}+1 - q + q^2$, and the hyperbolic volume is approximately $2.02988$.
Image taken from Wikimedia Commons.}
\begin{figure}[h]
\centering
\includegraphics[scale=.9]{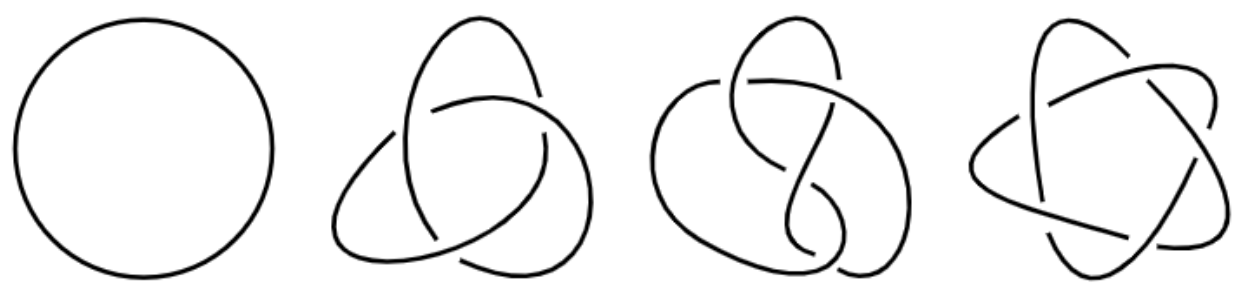}
 \caption{\one}
 \label{fig:knots}
\end{figure}
Knot invariants, which distinguish knots from each other, are independent of how a knot is drawn on the plane (the knot diagram).
Determining relationships between these invariant quantities is a central theme of knot theory.
See Appendix~\ref{sec:knot-invariants} for a brief overview of the invariants discussed in this work.
Perhaps the most famous invariant of a knot $K$ is the Jones polynomial $J_K(q)$, which is a Laurent polynomial with integer coefficients.
The original definition of the Jones polynomial was combinatorial~\cite{Jones1987}, but an intrinsically geometric definition and generalization was discovered soon thereafter~\cite{Witten1989}.
The generalizations found in~\cite{Witten1989} are known as ``colored" Jones polynomials and represent a family of knot invariants $J_{K,N}(q)$ labeled by a positive integer $N$ called the color.
The special value $N=2$, corresponding to a Wilson loop in the fundamental representation, recovers the Jones polynomial.
While the Jones polynomial is defined for any knot, some invariants exist only for subsets of knots.
An example of such an invariant is the hyperbolic volume of the knot's complement, denoted $\text{Vol}(K)$.
It is defined only if the manifold obtained by drilling out the knot from its $3$-dimensional ambient space admits a complete hyperbolic structure.
The vast majority of knots are hyperbolic~\cite{Thurston1982}, and we will restrict our attention to this case.
An important open problem in knot theory is to establish a conjecture that relates $J_{K,N}(q)$ to $\text{Vol}(K)$.
The volume conjecture~\cite{Kashaev1997,Murakami2001,Gukov2005} asserts that
\begin{equation}
\lim_{N \to \infty} \frac{2\pi \log |J_{K,N}(e^{\frac{2\pi i}{N}}) |}{N} = \text{Vol}(K) ~.
\label{eq:vc}
\end{equation}
The main idea of the volume conjecture is that the colored Jones polynomial in the large color limit contains information about the volume of $K$.

One might wonder if this property of the colored Jones polynomials extends to the original Jones polynomial.
Evaluating the Jones polynomial at $q=-1$, there is a surprising approximately linear relationship between $\log |J_K(-1)|$ and $\text{Vol}(K)$, but this correlation seems to apply only to a particular class of knots~\cite{Dunfield2000}.
Additionally, the so-called ``volume-ish" theorem~\cite{Dasbach2007} gives upper and lower bounds on $\text{Vol}(K)$ in terms of certain coefficients appearing in $J_K(q)$.
An improved relationship is achieved~\cite{Khovanov2003} by replacing $J_K(-1)$ with the reduced rank of the Khovanov homology, a homology theory $\mathcal{H}_{K}$ whose graded Euler characteristic is $J_K(q)$.
The cost of this improvement is that the Khovanov homology is a much more refined invariant of $K$ than the Jones polynomial, and one needs to work much harder to compute it~\cite{Khovanov2000}.
The most optimistic interpretation of these results is that there is a nonlinear relation $A$ mapping $J_K(q)$ to $\text{Vol}(K)$ along the lines of
Eq.~\eref{eq:vc},
but perhaps not quite as simple to write.
In this report, we provide evidence for this idea by directly estimating $A$ using a simple two hidden layer fully connected neural network.

\section{Setup and Result}
A neural network is a function which is constructed by training on several examples.
Suppose that we have a dataset $D= \{J_1, J_2,\ldots, J_m\}$, and to every element of $D$, there is an associated element in another set $S$:
\begin{eqnarray}
A: \{J_1, J_2, \ldots, J_m\} & \mapsto & \{ v_1, v_2, \ldots, v_m \} \subset S ~.
\label{eq:ass}
\end{eqnarray}
In our case, the $J_i$ are the Jones polynomials of knots, and the $v_i$ are the volumes of those knots.\footnote{Actually, the map we want to consider is $\widetilde{A}: \{\text{knots}\} \to \{\text{volumes}\}$ from the database of knots to their volumes. This is because Eq.~\eqref{eq:ass} is not a unique association; it is possible to have several knots that have the same Jones polynomials but different volumes. However, we can think of $A$ as a filtered version of $\widetilde{A}$, where we only give the network limited information about the knots, namely their Jones polynomials.
This is discussed further in the last paragraph of Appendix~\ref{sec:knot-invariants} and around Eq.~\eqref{eq:SchForm} in Appendix~\ref{sec:netdetails}.}
A neural network $f_{\theta}$ is a function (with an \textit{a priori} chosen architecture) which is designed to approximate the associations $A$ efficiently; the subscript $\theta$ denotes the internal parameters, called weights and biases, on which the neural network depends.
In order for the network to learn $A$, we divide the dataset $D$ into two parts:
a training set, $T = \{ J_1, J_2, \ldots, J_n \}$ chosen at random from $D$, and its complement, $T^c = \{ J'_1, J'_2, \ldots, J'_{m-n} \}$.
The neural network is taught the associations on the training set by tuning the internal parameters $\theta$ to approximate $A$ as closely as possible on $T$.
In general, $f_\theta(J_i) \ne v_i$ without overfitting the data.
We must instead minimize a suitably chosen loss function that captures the difference between the two.
Finally, we assess the performance of the trained network by applying it to the unseen inputs $J'_i\in T^c$ and comparing $f_\theta(J'_i)$ to the true answers $v'_i = A(J'_i)$.
See Appendix~\ref{sec:neural-networks} for more details about neural networks and our particular architecture and implementation.

Neural networks of appropriate size can approximate any function~\cite{Cybenko1989} and in general are composed of layers which perform matrix multiplication, bias vector addition, and a nonlinear activation function $\sigma$ which acts element-wise on vectors.
After encoding the Jones polynomial $J_K(q)$ in a vector $\vec{J}_K$ consisting of the integer coefficients and the maximum and minimum degree of the polynomial, our network can schematically be written as
\begin{equation}
f_\theta(\vec{J}_K) = \sum_a \sigma \left( W^2_\theta\cdot \sigma( W^1_\theta \cdot \vec{J}_K + \vec{b}^1_\theta) + \vec{b}^2_\theta \right)^a  ~,
\label{eq:nn}
\end{equation}
where $W^j_\theta$ and $\vec{b}^j_\theta$ are the weight matrices and bias vectors, respectively, of the $j^{\text{th}}$ hidden layer and the summation simply adds up the components of the output vector.
The input layer is padded with zeroes so that the vectors are of uniform length.
In our case, the inputs are vectors of length $18$.
The hidden layers have $100$ neurons each, and the final output layer is a summation over the output of the second hidden layer.
In the language of
Eq.~\eref{eq:nn},
$W^1_\theta$ is a $100 \times 18$ matrix, $\vec{b}^1_\theta$ is a length $100$ vector, $W^2_\theta$ is a $100 \times 100$ matrix, and $\vec{b}^2_\theta$ is a length $100$ vector, all with variable entries that are determined by training the network on data.
For data, we use a table of Jones polynomials and hyperbolic volumes for $313,209$ knots obtained from the online databases Knot Atlas~\cite{KnotAtlas} and SnapPy~\cite{SnapPy}. This includes all hyperbolic knots in Knot Atlas up to $15$ crossings. We implement and train $f_\theta$ in Mathematica~11~\cite{Wolfram} using built in functions that are completely unoptimized for the problem under consideration.
The loss function is proportional to the squared error in volume, and parameters are adjusted by stochastic gradient descent.
We follow the usual protocol for neural network training~\cite{LeCun2015}:
the network is shown a set of training data, and loss function gradients on this set are used to adjust the network parameters $\theta$ via backpropagation.

With this architecture, our network performs significantly better than any previous method of estimating $\text{Vol}(K)$ from $J_K(q)$.
Its simplicity and robustness suggest the existence of an almost exact nonlinear relationship between the two invariants which is more complicated than
Eq.~\eref{eq:vc},
but not by much.

In
Figure~\ref{fig:mainresult},
we show a plot of the accuracy of our trained model compared with the volume-ish bounds and Khovanov homology rank methods.
\newcommand\two{Scatterplots of predicted volume versus actual volume for various prediction methods with dashed black lines denoting perfect prediction.
(a) Prediction for $111,521$ alternating knots using the volume-ish theorem.
The predicted volumes were obtained by selecting a random real number in the allowed range prescribed by~\cite{Dasbach2007}.
(b) Prediction for $196,011$ knots (the subset of knots for which the Khovanov homology rank was readily available).
The predicted volumes were obtained by fitting a linear function to the set of points defined by $(\log (\text{rank} (\mathcal{H}_{K}) - 1), \text{Vol}(K))$ and then applying that function to $\log (\text{rank}(\mathcal{H}_{K})-1)$.
(c) Prediction for all $313,209$ knots using the neural network $f_\theta$.
The predicted volumes were obtained by training $f_\theta$ on $10$\% of the data and then applying $f_\theta$ to all of the Jones polynomials.}
\begin{figure}[h]       
\centering
\includegraphics[scale=.45]{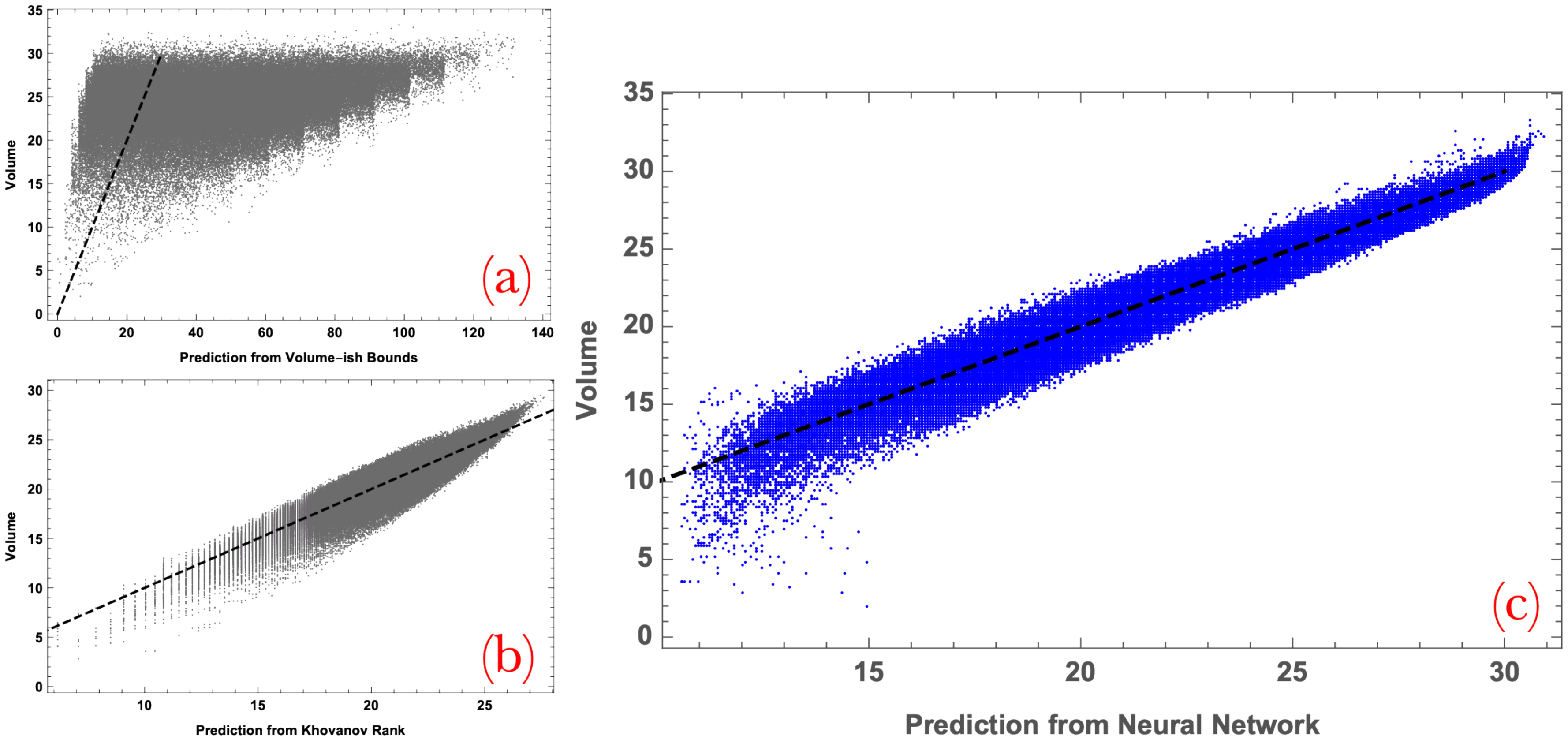}
 \caption{\two}
 \label{fig:mainresult}
\end{figure}
Random selection within the volume-ish bounds leads to an enormous error 
(Figure~\ref{fig:mainresult}a).
This is because the volume-ish bounds are fairly loose, and the range of allowed volumes is wide enough that large errors become unavoidable due to random selection.
The theorem applies only to knots for which crossings alternate between underhand and overhand, so we have restricted to this subset.
The Khovanov homology rank, on the other hand, applies more generally and can predict the volume with a mean error of approximately $4.67$\%
(Figure~\ref{fig:mainresult}b).
However, even the Khovanov homology rank predictions show a large spread around the perfect prediction line.
In
Figure~\ref{fig:mainresult}c,
we show our network's performance.
We compute the relative error
\begin{equation}
\delta f_\theta = \text{Mean} \left( \frac{|f_\theta(K') - \text{Vol}(K')|}{\text{Vol}(K')} \right) ~,
\label{eq:relerror}
\end{equation}
where $K'$ are knots belonging to the complement of the training set.
Averaging over $100$ runs, the relative error is $2.45\pm 0.10$\% when training on $10$\% of the data.
This error increases to $2.8$\% when training on just $1$\% of the data.
The neural network analysis applies to all knots in the database.
We notice that the spread between the prediction and the actual value decreases at larger volumes.
In part, this is because there is more data here as the number of possible knots and the mean volume both increase with crossing number.
\newcommand\three{The neural network quickly converges to optimal performance while training on a given fraction of the total dataset of $313,209$ knots.
Data points and the associated error bars are computed from averaging over $20$ trials each of which is trained on a randomly selected sample of the dataset.}
\begin{figure}
\centering
\fbox{\includegraphics[scale=.7]{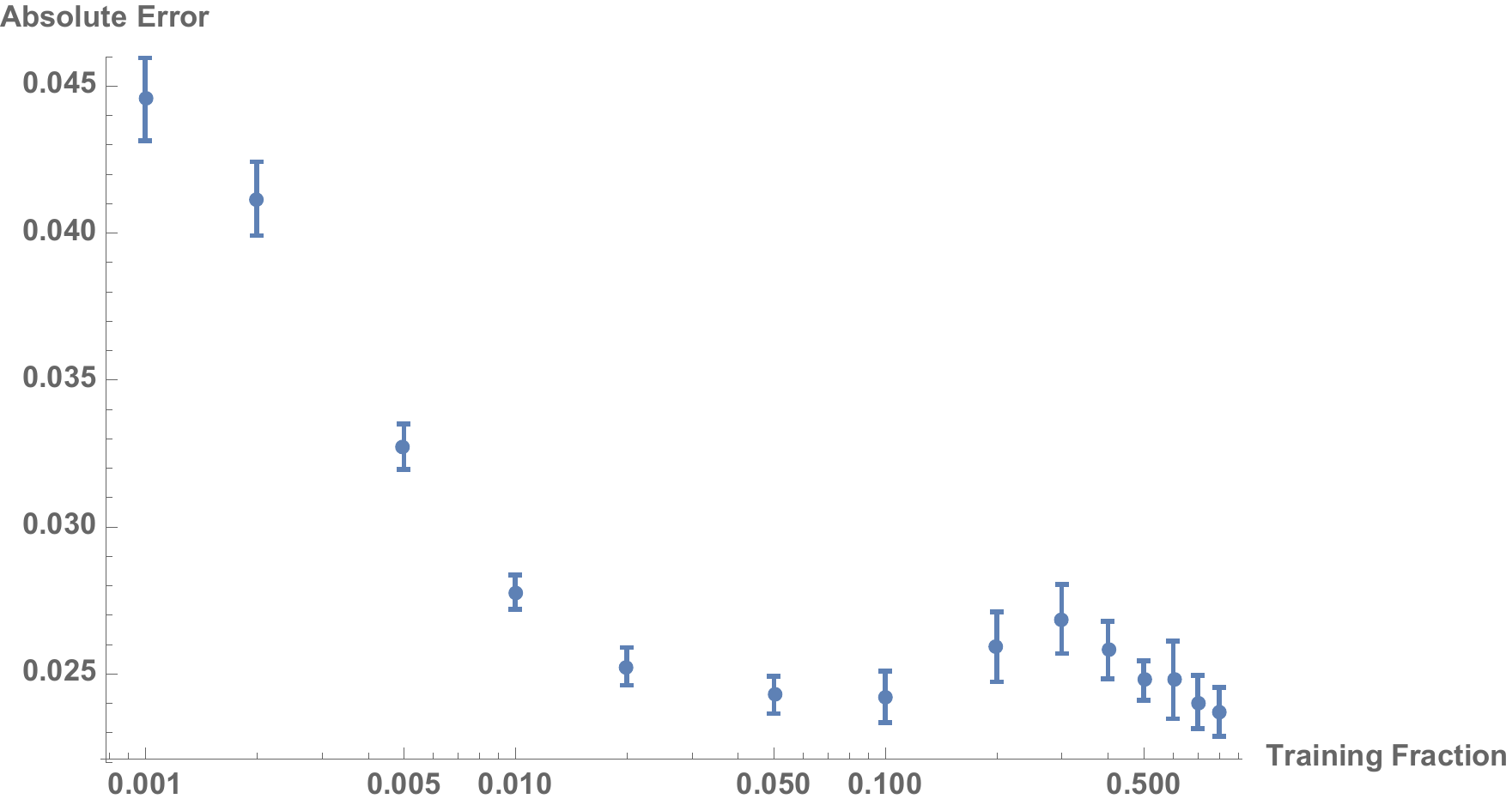}}
 \caption{\three}
 \label{fig:fracs}
\end{figure}
Figure~\ref{fig:fracs}
illustrates how little input is needed for the network to learn the correlation between the Jones polynomial and the hyperbolic volume: the horizontal axis gives the size of the training set as a fraction of the complete dataset, and the vertical axis gives the average relative error.
This is an instance of probably approximately correct learning~\cite{Valiant1984}.

Since it can extract very predictive features from small subsets of the data, this suggests $f_\theta$ is learning something fundamental that connects the Jones polynomial to the hyperbolic volume.
Indeed, $0.1$\% of the data is already enough to teach our network more (in terms of lower average error) about the hyperbolic volume than is known by the Khovanov homology rank function of~\cite{Khovanov2003}, despite the fact that $\mathcal{H}_{K}$ is a more refined knot invariant than $J_K(q)$, and therefore intuitively we would expect it contains more information about the volume.
Perhaps a neural network architecture which takes in aspects of the Khovanov homology as an input would perform even better in predicting the hyperbolic volume.

The performance of our very simple network is robust in the sense that adding extra layers, adding or removing a few neurons in each layer, changing the activation functions, and changing the loss function all have negligible effects on the resulting trained network accuracy.
Indeed, a single layer performs almost as well as the two layer architecture we have chosen.
The training of $f_\theta$ is relatively smooth and occurs quickly.
It can be accomplished on a laptop in under $3$ minutes.
We plot the loss versus the number of training rounds in Figure~\ref{fig:loss}.
\newcommand\four{Average loss versus number of training rounds for both training (orange curve) and test (blue curve) datasets.
The training set was 10\% of the data, chosen at random, and the test set was the complement.
The loss function can be viewed as a proxy for the error rate, and in our setup the two are proportional.}
\begin{figure}
\centering
\includegraphics[scale=.7]{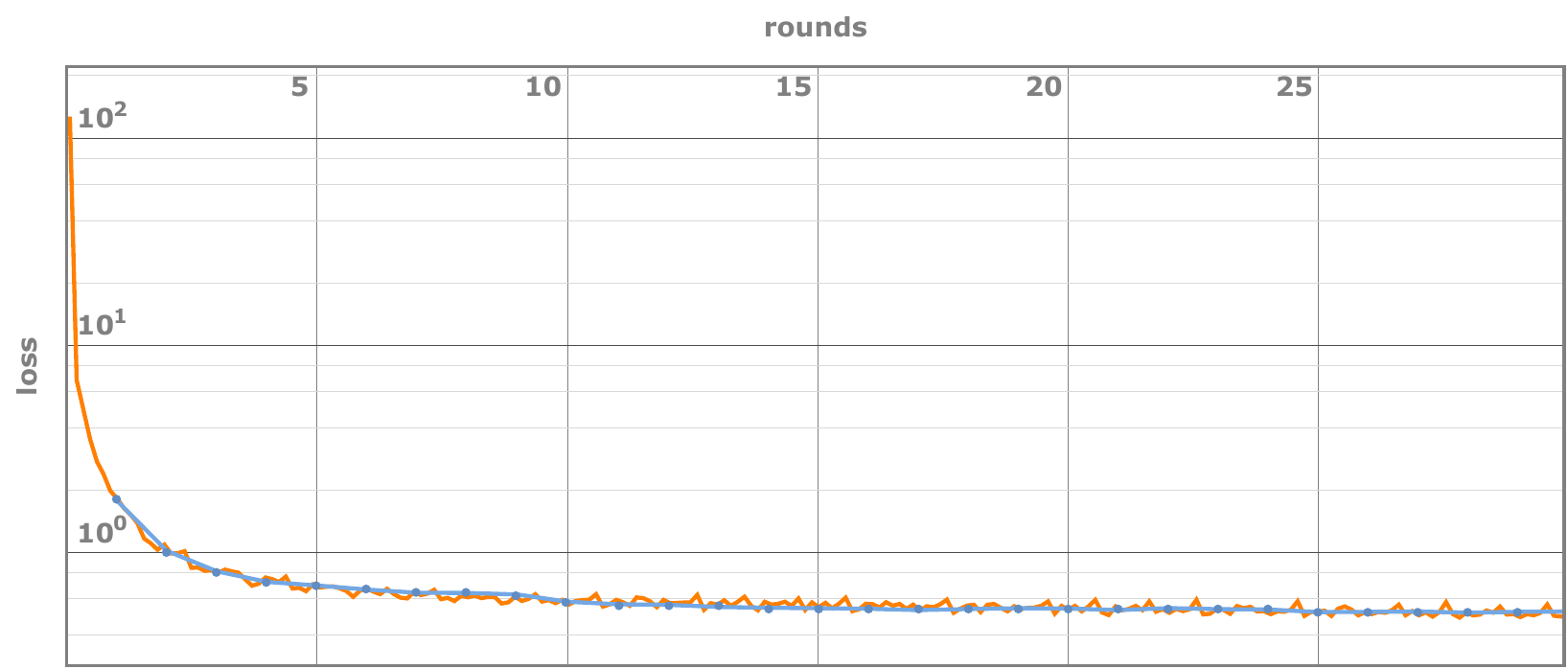}
 \caption{\four}
 \label{fig:loss}
\end{figure}
The neural network learns how to predict the hyperbolic volume from the Jones polynomial quickly, and most of the later rounds contribute only marginal improvements to the error rate.
Furthermore,
Figure~\ref{fig:loss}
shows the training and test sets are approximately equal in loss for essentially the entire duration of training despite the fact that the network never uses the test set for training.
This feature persists for much smaller training set sizes ($1$\%, $5$\% of data).
These observations support our conclusion from the error rate discussion and again suggest that the network can learn robust features after seeing just a small amount of data for a short amount of time.

The training data must be representative, however.
The volume of the knot complement serves as a proxy for the complexity of a knot.
Training only on the $25$\% of knots with the smallest volume, the neural network underpredicts the volumes of the remaining knots.
The error is $12.8$\%.
Seeding the training set with a small sampling of the higher volume knots restores the performance of the network.
See Appendix~\ref{sec:other-experiments} for other experiments we performed.

\section{Discussion}
We have shown that a relationship between the (uncolored) Jones polynomial $J_K(q)$ and the hyperbolic volume $\text{Vol}(K)$ similar in spirit to
the volume conjecture Eq.~\eref{eq:vc}
can be learned quickly and robustly using a deep neural network with only two hidden layers.
We now comment on some implications of our findings for knot theory and theoretical physics as well as potential directions for future work.
Perhaps the most obvious question is whether there is really a not-so-complicated function $A$ which exactly computes $\text{Vol}(K)$ from $J_K(q)$ with small corrections coming from other knot invariants.
There is some evidence suggesting that underlying the relationship between $J_K(q)$ and $\text{Vol}(K)$ is the theory of Khovanov homology~\cite{Khovanov2003}.
Recent work~\cite{Bull2018} shows that Hodge numbers of complete intersection Calabi--Yau threefolds can be computed by neural network classifiers and support vector machines in polynomial time offering a considerable simplification over traditional Gr\"obner basis methods, which are by comparison doubly exponential in time.
The Hodge numbers are dimensions of cohomology groups.
The existence of an underlying homology or cohomology theory could be a crucial aspect to machine learning this class of problems.

In theoretical physics, colored Jones polynomials appear as expectation values of Wilson loop operators in Chern--Simons theory~\cite{Witten1989}.
The volume conjecture has an interpretation~\cite{Gukov2005} in this context as a relationship between a double scaling limit of $SU(2)$ and the weak coupling limit of $SL(2,\mathbb{C})$ Chern--Simons theory.
In this report we demonstrate a potential connection between the \textit{strong} coupling limit of $SU(2)$ and the weak coupling limit of $SL(2,\mathbb{C})$ Chern--Simons theory.
Can other relationships between coupling regimes of topological quantum field theories be found using these neural network techniques to analyze expectation values?
The intimate association between knot invariants and Gromov--Witten invariants~\cite{Gopakumar1998,Ooguri1999} indicates that new insights about topological strings can also be gained by adapting machine learning techniques.
It might also be interesting to apply machine learning techniques to investigate the quantum entanglement structure of links studied in~\cite{Balasubramanian2016, Balasubramanian2018}.
Recently,~\cite{He2017,Krefl2017,Ruehle2017,Carifio2017} have pioneered investigations of the string landscape with machine learning techniques.
Exploring the mathematics landscape in a similar spirit, we expect that the strategy we employ of analyzing correlations between properties of basic objects can suggest new relationships of an approximate form.

\section*{Acknowledgements}
We are grateful to Dror Bar-Natan and Scott Morrison for correspondence and for computing HOMFLY polynomials at $15$ crossings.
We thank Vijay Balasubramanian, Robert de Mello Koch, Sergei Gukov, and Tassos Petkou for helpful conversations.
The work of VJ is supported by the South African Research Chairs Initiative of the Department of Science and Technology and the National Research Foundation.
The work of AK is supported by DoE grant DE-SC0013528. OP acknowledges support from the Simons Foundation (\#385592, VB) through the It From Qubit Simons
Collaboration, and the DoE contract FG02-05ER-41367.
All data used in this analysis is publicly available on the Knot Atlas~\cite{KnotAtlas} and SnapPy~\cite{SnapPy} websites.

\appendix 

\section{Overview of knot invariants}\label{sec:knot-invariants}
\begin{figure}[!h]
\centering
\includegraphics[scale=.67]{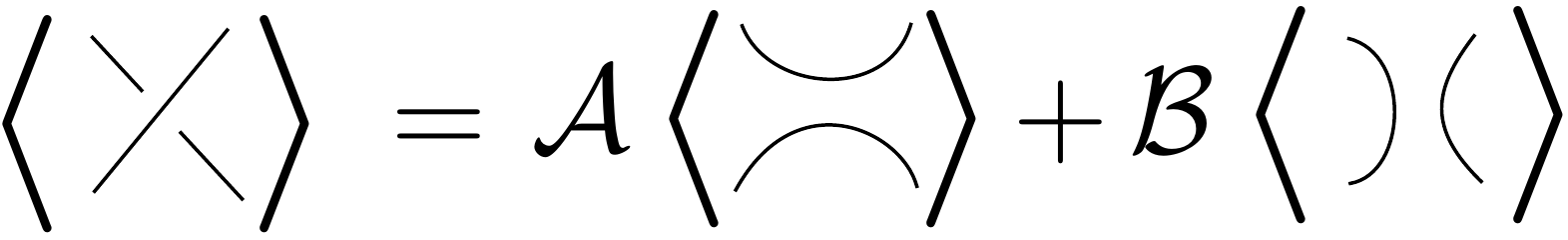}
\caption{\small{The smoothing relation in the definition of the Kauffman bracket. Each of the terms appearing on the right hand side refers to a choice of smoothing of the crossing on the left hand side.}\label{Kauffman}}
\end{figure}

In this section, we give a brief overview of the knot invariants of direct relevance to this work, namely the Jones polynomial and the hyperbolic volume (and also to some extent the Khovanov homology). All of these are topological invariants of a knot in the sense that they do not depend on a specific two-dimensional drawing of the knot, but depend only on its topology. Let us begin with the Jones polynomial. This is defined using the Kauffman bracket $\langle K \rangle$, where $K$ is the knot in question. The Kauffman bracket satisfies three conditions: (1) $\langle \emptyset \rangle = 1$, (2) $\langle \bigcirc \, K \rangle = -({\cal A}^2+{\cal B}^2) \langle K \rangle$ where $\bigcirc$ is the unknot, ${\cal A} = q^{1/4}$ and ${\cal B} = q^{-1/4}$, and (3) the smoothing relation shown in Figure~\ref{Kauffman}. 
These rules allow us to uniquely associate a Laurent polynomial in $q$ to every smoothing of the knot, and the sum of all these terms (\textit{i.e.}, over all the smoothings) is the Kauffman bracket (see~\cite{Kauffman1987} for details). The Jones polynomial is then equal to the Kauffman bracket up to an overall normalization constant:
\begin{equation}
J_K(q) = (-q^{3/4})^{w(K)} \frac{\langle K \rangle}{\langle \bigcirc \rangle},
\end{equation}
where $w(K)$ is the writhe of $K$, the number of overhand crossings minus the number of underhand crossings. 
 It was famously shown by Witten~\cite{Witten1989} that the Jones polynomial of a knot $K$ can also be thought of as the expectation value of a Wilson loop operator along $K$ in $SU(2)$ Chern--Simons theory. Since Chern--Simons theory is a (three-dimensional) topological quantum field theory, this gives a manifestly three-dimensional perspective for why the Jones polynomial is a topological invariant of the knot. Interestingly, the Jones polynomial also turns out to be a polynomial (in powers of $q$) with \emph{integer} coefficients. This fact was later explained by Khovanov homology. Very briefly, the Khovanov homology can be thought of as a categorification of the Jones polynomial. In Khovanov homology, one defines a Khovanov bracket in analogy with the Kauffman bracket, but we associate a tensor power of a graded vector space with every smoothing of the knot. By taking certain direct sums of these vector spaces and defining a suitable differential operator between them, we build a chain complex. It can then be shown that the Jones polynomial is the graded Euler characteristic of this complex, and thus the coefficients of the Jones polynomial are the dimensions of the vector spaces which appear in the chain complex. For more details, see~\cite{Khovanov2000, BarNatan2002}.

The other knot invariant which is central in this work is the hyperbolic volume of a knot. For any knot $K$ in $S^3$, the knot complement is defined as the manifold $M_K = S^3 - K$. More precisely, we remove a tubular neighborhood of the knot from $S^3$. Knots for which the knot complement admits a complete hyperbolic structure are called hyperbolic knots. For such a knot $K$, the complete hyperbolic structure on the knot complement $M_K$ is unique, and the corresponding volume of $M_K$ is called the hyperbolic volume $\text{Vol}(K)$ of the knot. The standard way to compute the hyperbolic volume (following~\cite{Thurston1982}) is to find a tetrahedral decomposition of the knot complement. Each tetrahedron can then be embedded in hyperbolic space, up to the specification of one complex number, often called the shape parameter of the tetrahedron. Requiring that all the tetrahedra in the knot complement fit together without any curvature singularities gives a set of algebraic constraints on the shape parameters, which can then be solved to obtain the shape parameters, and thus the desired hyperbolic structure. 
The volume of the knot complement is then the sum of the volumes of the individual tetrahedra.

The Jones polynomial by itself is not sufficient to identify a knot uniquely. For example, the knots $4_1$ (the figure-eight knot) and K$11$n$19$ have the same Jones polynomials but different volumes (the converse can also occur). There are $174,619$ unique Jones polynomials in our dataset.
 
\section{Neural networks}\label{sec:neural-networks}
Our aim is to construct a function $f_{\theta}$ which approximates the relation
\begin{equation}
A:  \{ J_K(q)\} \mapsto \left\{ \text{Vol}(K)  \right\}
\end{equation}
as closely as possible. We use a deep neural network to achieve this. A neural network $f_{\theta}$ is a (generally nonlinear) map from an input data vector $\vec{v}_{\text{in}} \in D$ to an output data vector $\vec{v}_{\text{out}}\in S$, where $\theta$ labels the internal parameters which the map involves. In our case, the input vectors are the Jones polynomials of the knots in our database, while the outputs are their corresponding volumes (so $S  = \mathbb{R})$. We divide the input vectors $D$ into the training set $T$ and its complement $T^c$. Given the relation $A: T \to S$ on the training set, the idea is to tune the parameters $\theta$ in such a way that $f_{\theta}$ reproduces $A$ on the training dataset as closely as possible. This is typically accomplished by picking some loss function $h(\theta)$, such as $h(\theta) = \sum_{i} || f_{\theta}(\vec{v}^{(i)}_{\text{in}}) -A(\vec{v}^{(i)}_{\text{in}}) ||^2$, where the sum is over $\vec{v}_{\text{in}}^{(i)} \in T$, and then minimizing $h(\theta)$ in the space of the parameters to find the point in parameter space at which the loss function is minimized. Having done so, we then apply the function $f_{\theta}$ to the set $T^c$ (which is so far unseen by the neural network) to test how well it approximates $A$ on it --- this ensures that $f_{\theta}$ is not trivially overfitting the data. It is known that neural networks of appropriate size can approximate any function~\cite{Cybenko1989}.

\begin{figure}[!h]
\centering
\includegraphics[scale=.7]{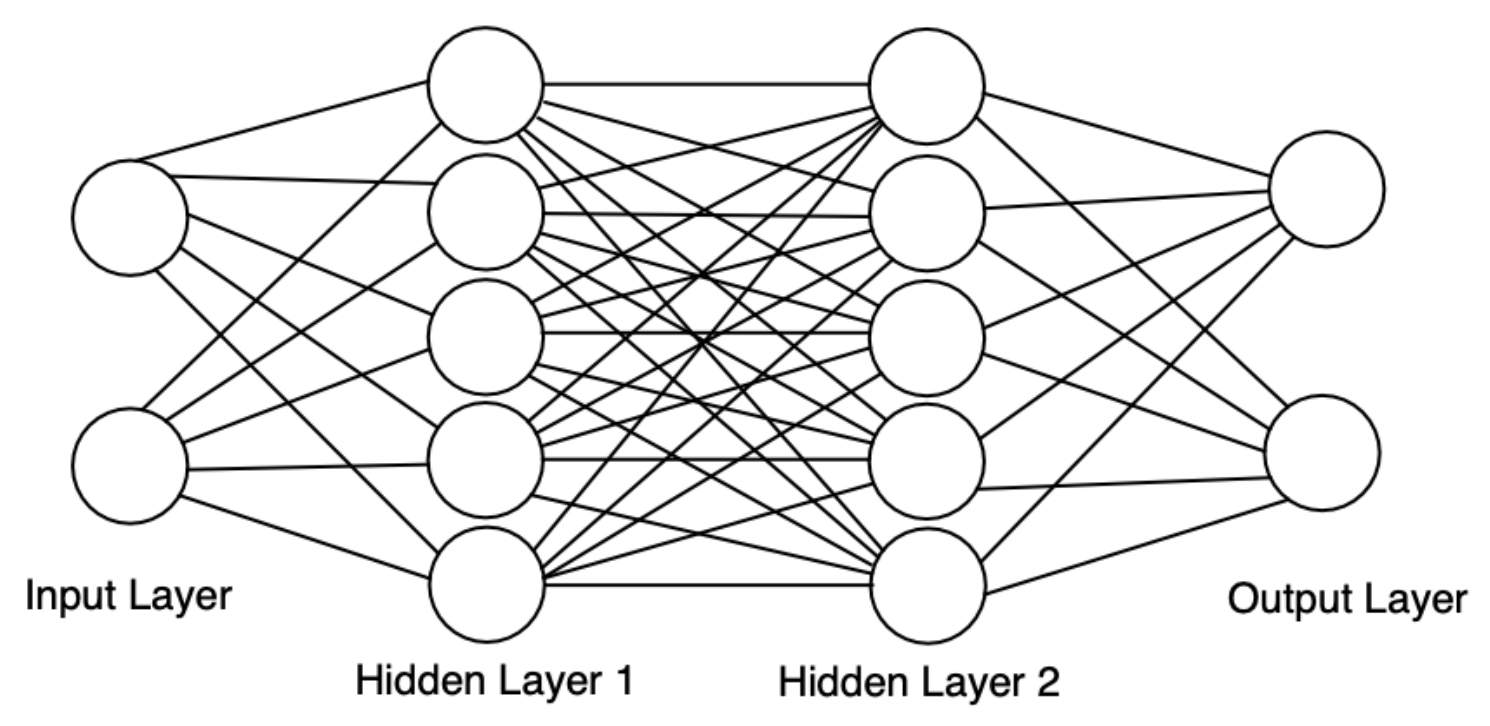}
\caption{An example of a two hidden layer fully connected neural network architecture.  Each hidden layer is shorthand for a matrix multiplication followed by a bias vector addition followed by an element-wise activation function; in our network, we use the logistic sigmoid function.  The final layer simply sums the components of the second hidden layer's output.  Our network $f_\theta$ takes an input vector of size $18$, has two $100$ neuron hidden layers, and a final summation output layer.\label{net}}
\end{figure}

Several interesting architectures of neural networks have been studied, but a simple architecture which will suffice for our purposes is the fully connected network (see Figure~\ref{net}). In this architecture, the network is composed of hidden layers which perform matrix multiplication and bias vector addition followed by element-wise application of an activation function $\sigma$.  
The network can thus be schematically written as
\begin{equation}
f_\theta(\vec{v}_{\text{in}}) =  L^n_{\theta}\left( \sigma\left( \cdots L^2_{\theta}\left(\sigma \left( L^1_{\theta}(\vec{v}_{\text{in}})\right)\right) \cdots\right)\right),\;\;\; L^{m}_{\theta}(\vec{v}) = W^{m}_{\theta} \cdot \vec{v} + \vec{b}^{m}_{\theta}  ,
\end{equation}
where $W^{m}_\theta$ and $\vec{b}^{m}_\theta$ are the weight matrices and bias vectors (respectively) of the $m^{\text{th}}$ hidden layer, and 
\begin{equation}
\sigma(\vec{v})^a = \sigma(\vec{v}^a),
\end{equation}
with $a$ being the vector index on the appropriate internal state. As stated previously, the idea is then to minimize the loss function on the training data by appropriately tuning the parameters $W^{m}_\theta$ and $\vec{b}^{m}_\theta$. This is achieved by using the backpropagation algorithm, which computes gradients of the loss function for each training data point and adjusts the parameters layer by layer in the network.

\subsection{Details of the network}\label{sec:netdetails}
As mentioned in the main text, the particular network we used in order to study the hyperbolic volume is of the form
\begin{equation}
f_\theta(\vec{J}_K) = \sum_a \sigma \left( W^2_\theta\cdot \sigma( W^1_\theta\cdot \vec{J}_K + \vec{b}^1_\theta) + \vec{b}^2_\theta \right)^a  ,
\end{equation}
where $\vec{J}_K = \left(p_{\text{min}}, p_{\text{max}}, c_1,c_2, \cdots, c_{\ell}\right)$ is a vector representation of the Jones polynomial
\begin{equation}
J_K(q) = \sum_{m = p_{\text{min}}}^{p_{\text{max}}} c_{m- p_{\text{min}}+1} q^{m}.
\end{equation}
Note that $\ell = 16$ is the maximum length of a Jones polynomial in our database. The coefficients $c_{ m> (p_{\text{max}} - p_{\text{min}}+1)}$ are simply set to zero in the vector representation. Further, $W^1_\theta$ is a $100 \times 18$ matrix, $\vec{b}^1_\theta$ is a $100$-vector, $W^2_\theta$ is a $100 \times 100$ matrix, and $\vec{b}^2_\theta$ is a $100$-vector. The activation function $\sigma$ is a logistic sigmoid function: 
\begin{equation}
\sigma(x) = \frac{1}{1+e^{-x}} .
\end{equation}
The network can be straightforwardly implemented in Mathematica 11.3.0.0~\cite{Wolfram} with the command\footnote{
In Mathematica 12.0.0.0, the \texttt{DotPlusLayer} command is replaced by \texttt{LinearLayer}.}
{\footnotesize
\begin{eqnarray}
\texttt{KnotNet} &=& \texttt{NetChain[\{DotPlusLayer[100]}, \nonumber\\
 & & \texttt{ElementwiseLayer[LogisticSigmoid], DotPlusLayer[100],} \nonumber\\
 & & \texttt{ElementwiseLayer[LogisticSigmoid], SummationLayer[]\},} \nonumber\\
 & & \texttt{"Input" -> \{18\}]; } \nonumber
 \end{eqnarray}}
 As mentioned before, the specific values of these internal parameters were found by training the network on a portion of the dataset, which can be implemented in Mathematica by the command \texttt{NetTrain}. The loss function is simply the mean squared error between the predicted volume and the true volume of the training example. We then test the accuracy of our network by applying it to the unseen knots in $T^c$. 
 
We are being conservative in estimating the error of our trained network. This is because the dataset contains several instances of knots with the same Jones polynomials but different volumes, \emph{i.e.}, the association $A$ we seek to teach the network to reproduce is not a function. Therefore, it may be that the network is taught one of the values of the volumes for such a Jones polynomial and then tested on a different value. We can repeat our analysis by keeping only the set of unique Jones polynomials within the dataset; when a Jones polynomial corresponds to several knots with different volumes, we select the volume of a randomly selected knot among these. In performing this experiment, we find that the relative error is unchanged. This could imply that the volume has the schematic form
\begin{equation}\label{eq:SchForm}
v_i = f(J_i) + \text{small corrections},
\end{equation}
and the success of the network is due to it learning $f$ very well. An examination of knots with the same Jones polynomial shows that the volumes tend to cluster; they differ on average by $2.83\%$. This is consistent with Eq.~\eqref{eq:SchForm} above. The deviation is larger for knots with smaller volumes, which is also consistent with the spread in the network predictions for small volumes in Figure~2c. 
 
Instead of listing out the weight matrices $W^i_\theta$ and the biases $\vec{b}^i_\theta$, which is difficult and unilluminating because of their size, we will show some of their properties in the Figures~\ref{spec1},~\ref{spec2} and~\ref{bias2} below. Notably, from the small error bars on the spectra of the weight matrices, it is evident that these are not random matrices, but are certain specific matrices which are central to the relation between the hyperbolic volume and the Jones polynomial. On the other hand, the large error bars on the biases suggest that they might not play a crucial role in the network. 
The plots here correspond to training on $10$\% of the total dataset, but similar plots for other fractions have the same profile.
In particular, the largest eigenvalue by magnitude in Figure~\ref{spec2} is essentially unchanged.
We also trained $f_\theta$ using a rectified linear unit activation function, but this network performed noticeably worse than the logistic sigmoid network.
It would be interesting to understand why this occurred, since the rectified linear unit has become relatively standard in the wider machine learning community.
It may be that our learning problem does not suffer from the vanishing gradient problem which rectified linear units resolve.

 \begin{figure}[!h]
 \centering
 \begin{tabular}{c c c}
 \includegraphics[scale=.5]{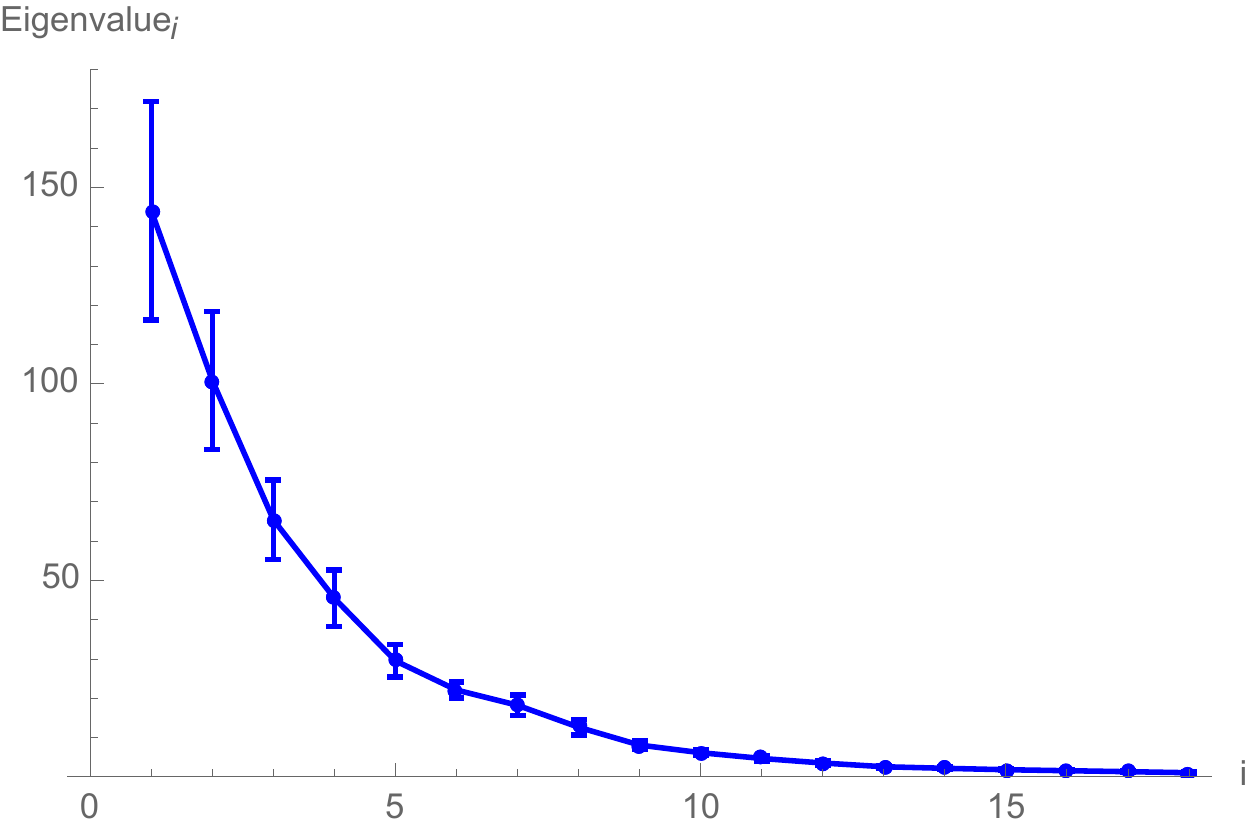} & \hspace{0.2cm}   & \includegraphics[scale=.5]{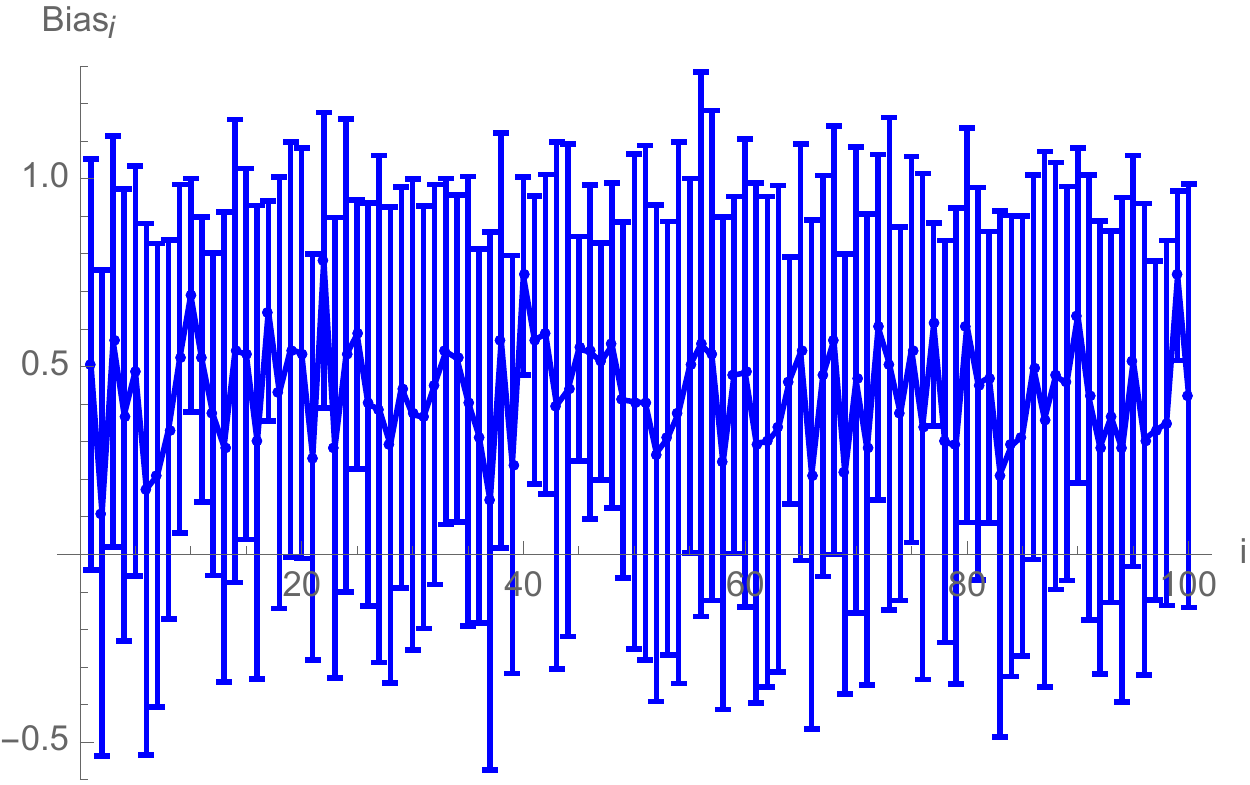}
 \end{tabular}
 \caption{\small{(Left) The eigenvalues of the matrix $(W^1_\theta)^TW^1_\theta$, where $W^1_\theta$ is the weight matrix of the first layer. The spectrum was averaged over $20$ runs at training fraction of 10\%, with the error bars marking the standard deviations. (Right) The biases $\vec{b}^1_\theta$ of the first layer averaged over $20$ runs, with the error bars marking the standard deviations. }\label{spec1}}
 \end{figure}

 \begin{figure}[!h]
 \centering
 \begin{tabular}{c c c}
 \includegraphics[scale=.4]{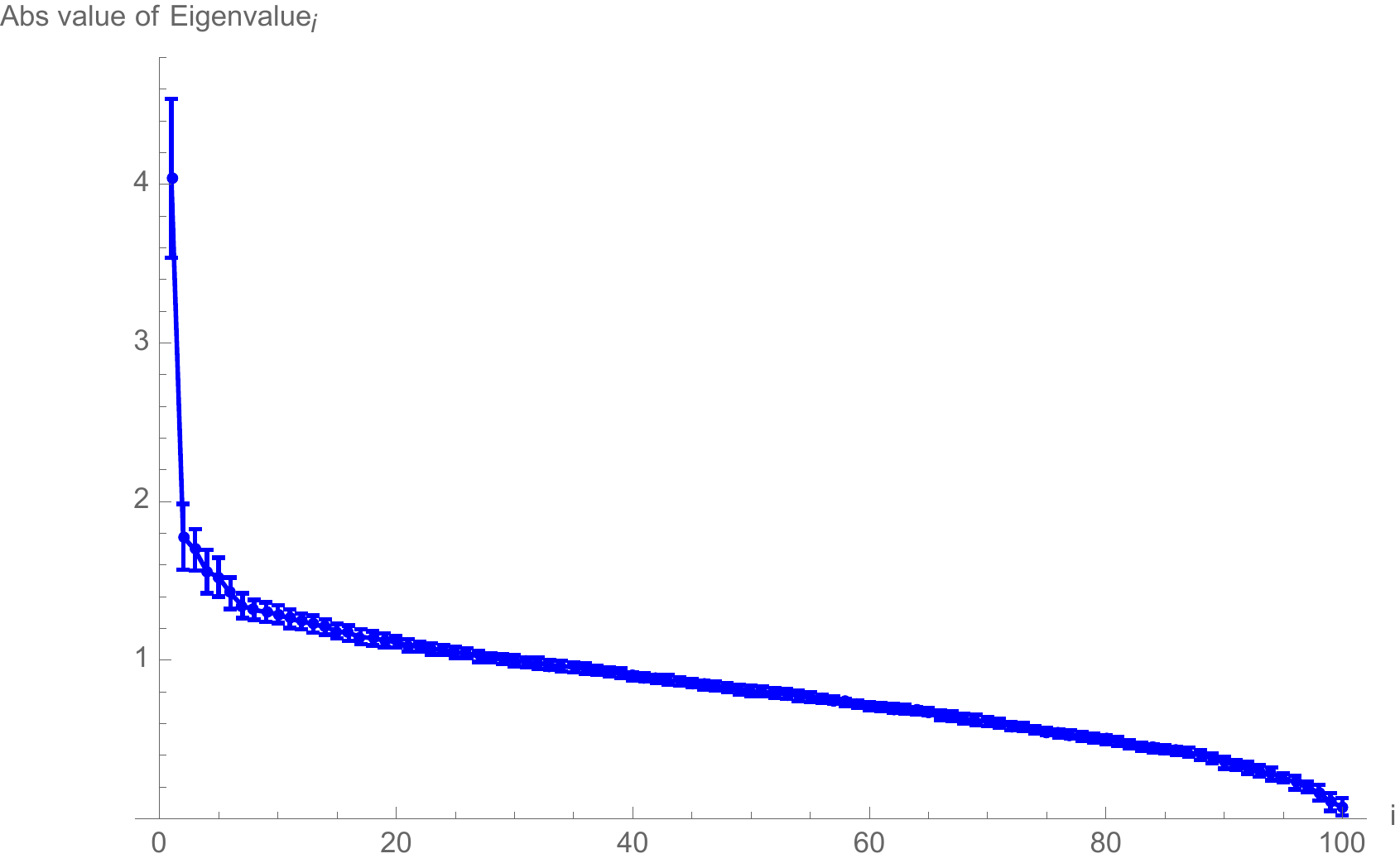} & \hspace{0.2cm}   &  \includegraphics[scale=.5]{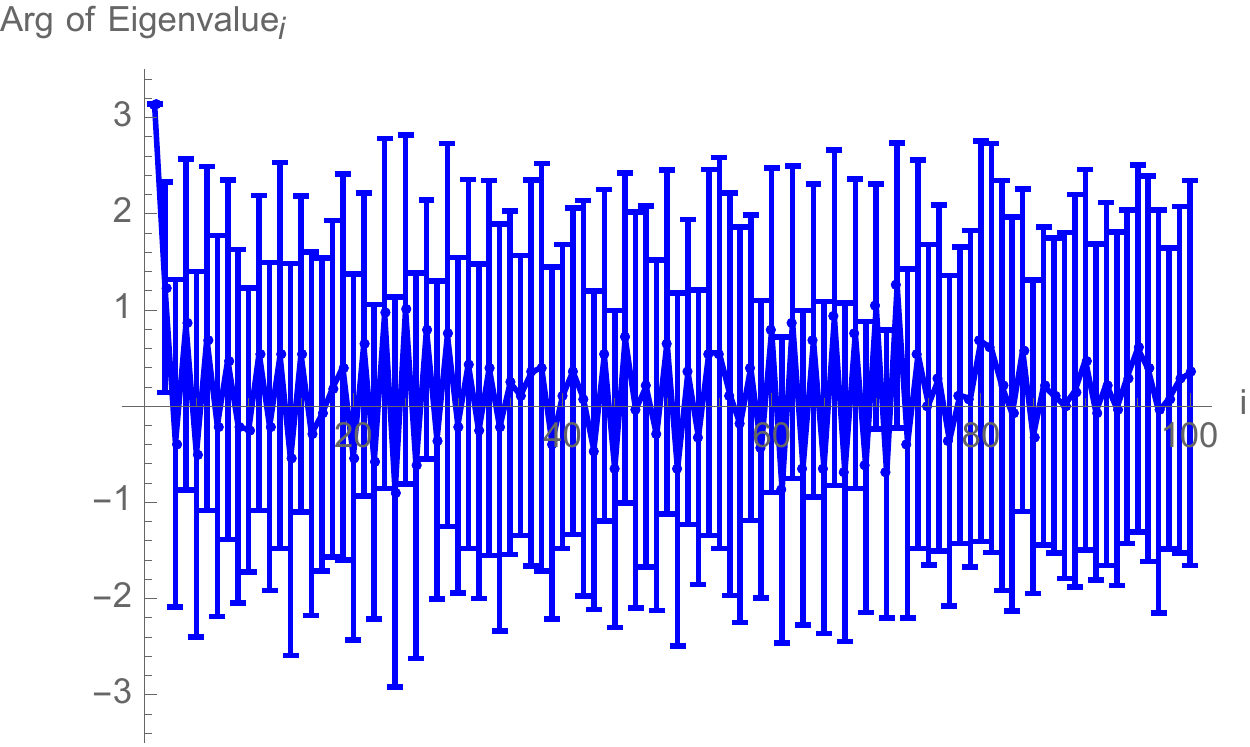} 
 \end{tabular}
 \caption{\small{(Left) The absolute values of the eigenvalues of the weight matrix $W^2_\theta$ of the second layer. The spectrum was averaged over $20$ runs, with the error bars marking the standard deviations. (Right) The phases of the eigenvalues of the weight matrix $W^2_\theta$. Note that the largest magnitude eigenvalue is always real and negative. This may be a consequence of a generalized version of the Perron--Frobenius theorem.}\label{spec2}}
 \end{figure}
  \begin{figure}[!h]
 \centering
 \includegraphics[scale=.5]{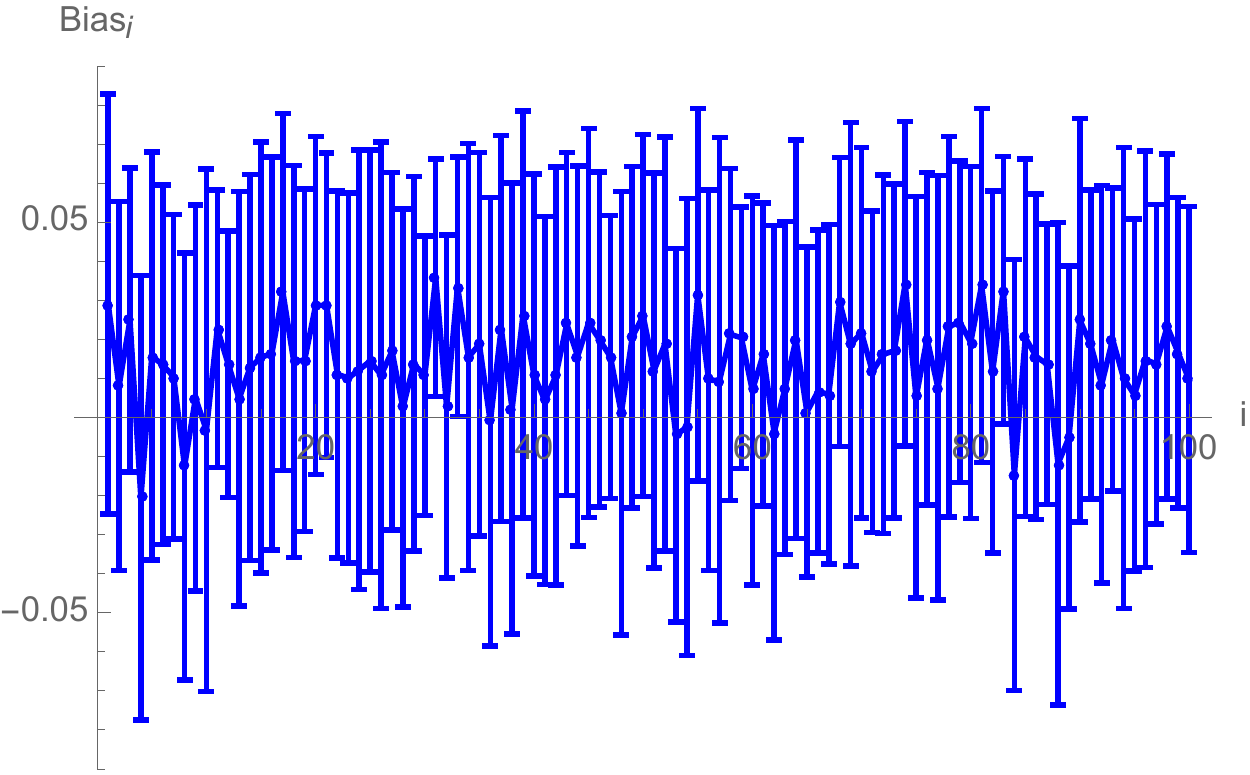}
 \caption{\small{The biases $\vec{b}^2_\theta$ of the second layer averaged over $20$ runs, with the error bars marking the standard deviations. }\label{bias2}}
 \end{figure}

\section{Other experiments}\label{sec:other-experiments}
Our results are unchanged using a larger dataset consisting of the $1,701,913$ hyperbolic knots with $16$ or fewer crossings.
Training on $2$\% of this set and averaging over $20$ trials, the absolute error is $2.40$\%.
Since the Mathematica code takes significantly longer to execute with the larger dataset, in order to improve our statistics, we have focused only on knots with up to $15$ crossings.
The training dataset necessary to ensure $97$\% accuracy requires $\sim 10,000$ examples for various crossing numbers.
This suggests that the performance of the neural network is very high with respect to the rate of growth of knots as a function of crossing number.
As noted in the letter, the training set should be representative of the full dataset.

There are many knot invariants known, and we tried to find other such relationships using similar techniques to those discussed in this paper.
These experiments had varying success.
We failed to reproduce the hyperbolic volume when training our network on the braid words, which capture all information about the knot in a compressed form.
We also failed to predict the Chern--Simons invariant (which is the imaginary part of the integral of the Chern--Simons three-form on the knot complement) from the Jones polynomial.
It appears on an equal footing with the volume in the generalized volume conjecture~\cite{Gukov2005}.
We succeeded up to $10\%$ error in reproducing the minimum and maximum degrees of the Jones polynomial from the braid word.
We attempted to learn the volume-ish theorem of~\cite{Dasbach2007} but the results were inconclusive.
We also attempted to learn a compressed form of the Jones polynomial from pictures of the knots using a convolutional neural network, but this did not work.
We did not have enough data to attempt any learning on the A-polynomial of~\cite{Cooper1994}, but it may be worth pursuing because it is more obviously connected to the hyperbolic volume.
The relationship between the Jones polynomial and volume is particularly striking in light of these failures.
Further investigation along these lines is warranted (see for instance~\cite{Hughes2016} for similar ideas).

\bibliography{scibib}

\providecommand{\href}[2]{#2}\begingroup\raggedright\begin{thebibliography}{10}

\bibitem{Witten1989}
E.~Witten, \emph{Quantum field theory and the {J}ones polynomial}, {\emph{Comm.
  Math. Phys.} {\bf 121} (1989) 351--399}.

\bibitem{Jones1990}
V.~F.~R. Jones, \emph{Knot theory and statistical mechanics}, {\emph{Scientific
  American} {\bf 263} (1990) 98--105}.

\bibitem{Sumners1995}
D.~W. Sumners, \emph{Lifting the curtain: {U}sing topology to probe the hidden
  action of enzymes}, {\emph{Notices of the AMS} {\bf 42} (1995) 528--537}.

\bibitem{Horner2016}
K.~E. Horner, M.~A. Miller, J.~W. Steed and P.~M. Sutcliffe, \emph{Knot theory
  in modern chemistry}, \href{http://dx.doi.org/10.1039/C6CS00448B}{\emph{Chem.
  Soc. Rev.} {\bf 45} (2016) 6432--6448}.

\bibitem{Ashley1944}
C.~Ashley, \emph{The Ashley Book of Knots}.
\newblock Doubleday, New York, 1944.

\bibitem{Jablan2012}
S.~Jablan, L.~Radović, R.~Sazdanović and A.~Zeković, \emph{Knots in art},
  {\emph{Symmetry} {\bf 4} (2012-6-05) 302--328}.

\bibitem{Jones1987}
V.~F.~R. Jones, \emph{Hecke algebra representations of braid groups and link
  polynomials}, \href{http://dx.doi.org/10.2307/1971403}{\emph{The Annals of
  Mathematics} {\bf 126} (Sep, 1987) 335}.

\bibitem{Thurston1982}
W.~P. Thurston, \emph{Three dimensional manifolds, {K}leinian groups and
  hyperbolic geometry}, {\emph{Bull. Amer. Math. Soc. (N.S.)} {\bf 6} (05,
  1982) 357--381}.

\bibitem{Kashaev1997}
R.~M. Kashaev, \emph{The hyperbolic volume of knots from quantum dilogarithm},
  \href{http://dx.doi.org/10.1023/a:1007364912784}{\emph{Letters in
  Mathematical Physics} {\bf 39} (1997) 269--275},
  [\href{https://arxiv.org/abs/arXiv:q-alg/9601025}{{\tt
  arXiv:q-alg/9601025}}].

\bibitem{Murakami2001}
H.~Murakami and J.~Murakami, \emph{The colored {J}ones polynomials and the
  simplicial volume of a knot},
  \href{http://dx.doi.org/10.1007/BF02392716}{\emph{Acta Math.} {\bf 186}
  (2001) 85--104}, [\href{https://arxiv.org/abs/arXiv:math/9905075}{{\tt
  arXiv:math/9905075}}].

\bibitem{Gukov2005}
S.~Gukov, \emph{Three-dimensional quantum gravity, {C}hern-{S}imons theory, and
  the {A}-polynomial},
  \href{http://dx.doi.org/10.1007/s00220-005-1312-y}{\emph{Comm. Math. Phys.}
  {\bf 255} (Mar, 2005) 577--627},
  [\href{https://arxiv.org/abs/arXiv:hep-th/0306165}{{\tt
  arXiv:hep-th/0306165}}].

\bibitem{Dunfield2000}
N.~Dunfield, \emph{An interesting relationship between the {J}ones polynomial
  and hyperbolic volume},
  {\emph{https://faculty.math.illinois.edu/$\sim$nmd/preprints/misc/dylan/index.html}
  (2000) }.

\bibitem{Dasbach2007}
O.~Dasbach and X.-S. Lin, \emph{A volumish theorem for the {J}ones polynomial
  of alternating knots},
  \href{http://dx.doi.org/10.2140/pjm.2007.231.279}{\emph{Pacific Journal of
  Mathematics} {\bf 231} (Jun, 2007) 279--291},
  [\href{https://arxiv.org/abs/arXiv:math/0403448}{{\tt arXiv:math/0403448}}].

\bibitem{Khovanov2003}
M.~Khovanov, \emph{Patterns in knot cohomology, {I}}, {\emph{Experiment. Math.}
  {\bf 12} (2003) 365--374},
  [\href{https://arxiv.org/abs/arXiv:math/0201306v1}{{\tt
  arXiv:math/0201306v1}}].

\bibitem{Khovanov2000}
M.~Khovanov, \emph{A categorification of the {J}ones polynomial},
  \href{http://dx.doi.org/10.1215/S0012-7094-00-10131-7}{\emph{Duke Math. J.}
  {\bf 101} (02, 2000) 359--426},
  [\href{https://arxiv.org/abs/arXiv:math/9908171}{{\tt arXiv:math/9908171}}].

\bibitem{Cybenko1989}
G.~Cybenko, \emph{Approximation by superpositions of a sigmoidal function},
  \href{http://dx.doi.org/10.1007/BF02551274}{\emph{Mathematics of Control,
  Signals and Systems} {\bf 2} (Dec, 1989) 303--314}.

\bibitem{KnotAtlas}
\emph{The knot atlas}, {\emph{http://katlas.org/} (2015) }.

\bibitem{SnapPy}
M.~Culler, N.~M. Dunfield, M.~Goerner and J.~R. Weeks, \emph{Snap{P}y, a
  computer program for studying the geometry and topology of $3$-manifolds},
  {\emph{http://snappy.computop.org/} (2018) }.

\bibitem{Wolfram}
\emph{Mathematica 11.3.0.0}.
\newblock Wolfram{\ }Research{\ }Inc., 2018.

\bibitem{LeCun2015}
Y.~LeCun, Y.~Bengio and G.~Hinton, \emph{Deep learning},
  \href{http://dx.doi.org/10.1038/nature14539}{\emph{Nature} {\bf 521} (May,
  2015) 436--444}.

\bibitem{Valiant1984}
L.~G. Valiant, \emph{A theory of the learnable}, {\emph{Commun. {ACM}} {\bf 27}
  (1984) 1134--1142}.

\bibitem{Bull2018}
K.~Bull, Y.-H. He, V.~Jejjala and C.~Mishra, \emph{{Machine learning CICY
  threefolds}},
  \href{http://dx.doi.org/10.1016/j.physletb.2018.08.008}{\emph{Phys. Lett.}
  {\bf B785} (2018) 65--72}, [\href{https://arxiv.org/abs/1806.03121}{{\tt
  1806.03121}}].

\bibitem{Gopakumar1998}
R.~Gopakumar and C.~Vafa, \emph{{On the gauge theory / geometry
  correspondence}},
  \href{http://dx.doi.org/10.4310/ATMP.1999.v3.n5.a5}{\emph{Adv. Theor. Math.
  Phys.} {\bf 3} (1999) 1415--1443},
  [\href{https://arxiv.org/abs/hep-th/9811131}{{\tt hep-th/9811131}}].

\bibitem{Ooguri1999}
H.~Ooguri and C.~Vafa, \emph{{Knot invariants and topological strings}},
  \href{http://dx.doi.org/10.1016/S0550-3213(00)00118-8}{\emph{Nucl. Phys.}
  {\bf B577} (2000) 419--438},
  [\href{https://arxiv.org/abs/hep-th/9912123}{{\tt hep-th/9912123}}].

\bibitem{Balasubramanian2016}
V.~Balasubramanian, J.~R. Fliss, R.~G. Leigh and O.~Parrikar,
  \emph{{Multi-boundary entanglement in {C}hern-{S}imons theory and link
  invariants}}, \href{http://dx.doi.org/10.1007/JHEP04(2017)061}{\emph{JHEP}
  {\bf 04} (2017) 061}, [\href{https://arxiv.org/abs/1611.05460}{{\tt
  1611.05460}}].

\bibitem{Balasubramanian2018}
V.~Balasubramanian, M.~DeCross, J.~Fliss, A.~Kar, R.~G. Leigh and O.~Parrikar,
  \emph{{Entanglement entropy and the colored {J}ones polynomial}},
  \href{http://dx.doi.org/10.1007/JHEP05(2018)038}{\emph{JHEP} {\bf 05} (2018)
  038}, [\href{https://arxiv.org/abs/1801.01131}{{\tt 1801.01131}}].

\bibitem{He2017}
Y.-H. He, \emph{{Machine-learning the string landscape}},
  \href{http://dx.doi.org/10.1016/j.physletb.2017.10.024}{\emph{Phys. Lett.}
  {\bf B774} (2017) 564--568}.

\bibitem{Krefl2017}
D.~Krefl and R.-K. Seong, \emph{{Machine learning of Calabi-Yau volumes}},
  \href{http://dx.doi.org/10.1103/PhysRevD.96.066014}{\emph{Phys. Rev.} {\bf
  D96} (2017) 066014}, [\href{https://arxiv.org/abs/1706.03346}{{\tt
  1706.03346}}].

\bibitem{Ruehle2017}
F.~Ruehle, \emph{{Evolving neural networks with genetic algorithms to study the
  string landscape}},
  \href{http://dx.doi.org/10.1007/JHEP08(2017)038}{\emph{JHEP} {\bf 08} (2017)
  038}, [\href{https://arxiv.org/abs/1706.07024}{{\tt 1706.07024}}].

\bibitem{Carifio2017}
J.~Carifio, J.~Halverson, D.~Krioukov and B.~D. Nelson, \emph{{Machine learning
  in the string landscape}},
  \href{http://dx.doi.org/10.1007/JHEP09(2017)157}{\emph{JHEP} {\bf 09} (2017)
  157}, [\href{https://arxiv.org/abs/1707.00655}{{\tt 1707.00655}}].

\bibitem{Kauffman1987}
L.~H. Kauffman, \emph{State models and the {J}ones polynomial},
  \href{http://dx.doi.org/https://doi.org/10.1016/0040-9383(87)90009-7}{\emph{Topology}
  {\bf 26} (1987) 395 -- 407}.

\bibitem{BarNatan2002}
D.~Bar-Natan, \emph{On {K}hovanov's categorification of the {J}ones
  polynomial}, \href{http://dx.doi.org/10.2140/agt.2002.2.337}{\emph{Algebr.
  Geom. Topol.} {\bf 2} (2002) 337--370},
  [\href{https://arxiv.org/abs/arXiv:math/0201043}{{\tt arXiv:math/0201043}}].

\bibitem{Cooper1994}
D.~Cooper, M.~Culler, H.~Gillet, D.~D.~Long and P.~Shalen, \emph{Plane curves
  associated to character varieties of 3-manifolds},
  \href{http://dx.doi.org/10.1007/BF01231526}{\emph{Inventiones Mathematicae}
  {\bf 118} (01, 1994) 47--84}.

\bibitem{Hughes2016}
M.~C. Hughes, \emph{A neural network approach to predicting and computing knot
  invariants},  \href{https://arxiv.org/abs/arXiv:1610.05744}{{\tt
  arXiv:1610.05744}}.

\end{thebibliography}\endgroup

\bibliographystyle{JHEP}

\end{document}